\documentclass[natbib]{svjour3}                     
\smartqed  
\usepackage{graphicx}
\usepackage{amssymb}
\newcommand\vv{{\bf v}}
\newcommand\fv{{\bf f}}
\newcommand\uv{{\bf u}}
\newcommand\Ev{{\bf E}}
\newcommand\Bv{{\bf B}}
\begin{document}

\title{Diffusive shock acceleration and magnetic field amplification 
}

\titlerunning{MFA and DSA}        

\author{K.~M.~Schure         \and
        A.~R.~Bell \and L.~O'C Drury \and A.~M.~Bykov  
}


\institute{K.M. Schure \at
              Department of Physics, University of Oxford, Clarendon Laboratory, Parks Road, Oxford OX1 3PU, United Kingdom \\
              \email{k.schure1@physics.ox.ac.uk}           
           \and
           A.R. Bell \at
               Department of Physics, University of Oxford, Clarendon Laboratory, Parks Road, Oxford OX1 3PU, United Kingdom \\
              \email{t.bell1@physics.ox.ac.uk}      
               \and
           A.M. Bykov\at
               Ioffe Institute for Physics and Technology, 19402 St. Petersburg, Russia \\
              \email{byk@astro.ioffe.ru}     
                \and
           L. O'C. Drury \at
               Dublin Institute for Advanced Studies, 31 Fitzwilliam Place, Dublin 2, Ireland  \\
              \email{ld@cp.dias.ie} 
}

\date{Received: date / Accepted: date}

\maketitle

\begin{abstract}
Diffusive shock acceleration is the theory of particle acceleration through multiple shock crossings. In order for this process to proceed at a rate that can be reconciled with observations of high-energy electrons in the vicinity of the shock, and for cosmic rays protons to be accelerated to energies up to observed galactic values, significant magnetic field amplification is required. In this review we will discuss various theories on how magnetic field amplification can proceed in the presence of a cosmic ray population. On both short and long scales, cosmic ray streaming can induce instabilities that act to amplify the magnetic field. Developments in this area that have occurred over the past decade are the main focus of this paper.
\end{abstract}

\section{Introduction}
\label{sec:intro}

The theory of diffusive shock acceleration (DSA) originates from the idea originally posed by \citet{1949Fermi} that cosmic rays are scattered by waves to be isotropic in their local frame. Scattering was proposed to happen in clouds where a turbulent magnetic field would isotropise the cosmic rays. Since on average head-on collisions are more frequent, a net acceleration occurs that naturally creates a powerlaw. Diffusive shock acceleration, proposed by various authors in the late seventies \citep{1977Axfordetal, 1977Krymskii, 1978Bella, 1978Bellb, 1978BlandfordOstriker} is the discovery that this acceleration proceeds much faster in the vicinity of a shock. When crossing the shock, the first collision is always head on, thus allowing the acceleration to proceed at a significantly faster rate, making it an attractive mechanism to accelerate cosmic rays to high energies. The faster the shock velocity, the larger the energy gains upon transition of the shock. The theory predicts a powerlaw with a spectral index that matches quite closely the observed powerlaw of cosmic rays arriving on Earth.

The first argument that pointed to supernova remnants (SNRs) as the prime sources of Galactic cosmic rays was based on the energy budget \citep{1934BaadeZwicky}. The energy required to replenish the cosmic rays against their losses from the Galaxy amounts to about 10\% of that available in SNRs. However, it was not until the late seventies that a theory on how to transfer kinetic energy efficiently into the acceleration of cosmic rays had been developed. As shown in this review, even though the basic test particle theory is simple, the intrinsic nonlinearity of this process makes it a difficult one to grasp in full and more work needs to be done to understand the process from start to finish.

The strong support that DSA can work as the major mechanism to accelerate cosmic rays came from observations of the thin X-ray rims at the blast waves of SNRs. The magnetic field required to confine the cosmic rays in the vicinity of the shock is much higher than the mean interstellar magnetic field \citep{1994Achterbergetal, 2003VinkLaming,2005Voelketal}. Originally it was proposed that the magnetic field could be amplified resonantly \citep{1967Lerche, 1969KulsrudPearce, 1974Wentzel, 1975Skilling}, but it is not clear that the magnetic field will be able to grow beyond $\delta B/B_0 \approx 1$ when the resonance condition is lost, as would be required in order to explain the observations. In the past decade, a number of theories has arisen that could potentially explain amplification of the magnetic field to values corresponding to those observed in supernova remnants.  

However, there is no reason to believe that only SNR shocks accelerate particles. Shocks are abundant in the universe on all scales, and how and when they become efficient accelerators is an active area of current research. Locally, cosmic rays are accelerated in heliospheric plasmas. On larger scales, evidence exists for accelerated electrons in the lobes of radio galaxies, as suggested early on by \citet{1974BlandfordRees} and backed up by more detailed observations and modelling \citep[e.g.][]{1991Carillietal,2009Crostonetal,2011BlundellFabian}. Furthermore clusters of galaxies have been observed to contain nonthermal radio emission suggesting active particle acceleration \citet{2006Bagchietal,2008Ferrarietal}.

Support for DSA on much larger scales has only recently been discovered with detailed and spectral observations in radio wavelengths of Megaparsec scale shocks \citep{2010vanWeerenetal}. The spectral hardening downstream from the supposed shock front in these systems indicates that active acceleration proceeds at the shock front itself. The shocks are believed to be the result of mergers of clusters of galaxies.  The magnetic field deduced from the radio observations indicate a field strength of the order of $\mu$G, which is much higher than would normally be expected in the intra-cluster medium \citep{2011Bruggenetal}. This is an additional indication the the process of DSA and magnetic field amplification is intrinsically linked and occurs at shocks on all scales. Because of the scales of these shocks potentially protons can be accelerated to energies of $10^{19}$~eV, although no direct evidence has been found yet. Magnetic field amplification through streaming cosmic rays has also been suggested as a source for primordial magnetic fields~\citep{2011MiniatiBell}.

On the Galactic scale, in addition to SNRs, shocks around superbubbles have been discussed as accelerators \citep[e.g.][]{1992BykovFleishman,2001BykovToptygin,2004Parizotetal,2008Binnsetal,2009Butt,2010FerrandMarcowith}. Superbubbles are formed as a result of cumulative outflows from an assembly of massive stars, possibly enforced by the supernova explosions themselves. The outer region thus formed is a shock of large proportions that could potentially be a cosmic ray accelerator, or various multiple shocks can consecutively act to accelerate particles, which may modify the resulting spectral index.

Although in reality many deviations arise due to e.g. nonlinear modification of the shock structure, magnetic field obliquity, geometric effects, time-dependence, and magnetic field amplification, the basic theory still holds. However, in order to be able to compare the theory with observations, all of the complicating factors need to be taken into account. In this review, we focus on the advances in theory that have been made over the past decade. More specifically, we will discuss how the amplified magnetic field required for efficient acceleration is a direct result of DSA and the presence of cosmic rays. Details of the other processes can be found in earlier review articles \citep{1983Drury, 1987BlandfordEichler,2001MalkovDrury,2005Hillas}.

In Section~\ref{sec:dsa} we will briefly review the original theory of diffusive shock acceleration and demonstrate how the process naturally results in a power law cosmic ray spectrum. In Sections~\ref{sec:mfa}--\ref{sec:longwavelength} we will discuss a number of theories that couple diffusive shock acceleration to magnetic field amplification. We will mainly focus on the most recent theories. In Section~\ref{sec:nonideal} we will discuss possible deviations from the source spectrum as a result of shock obliquity, nonlinearity, and time-dependence. We refer to other chapters in this book for a treatments of DSA in relativistic shocks, and to the observational insights and developments.

\section{Diffusive shock acceleration}
\label{sec:dsa}

The powerlaw in energy that results from diffusive shock acceleration can be understood in different ways, highlighted by different authors in its period of discovery. A crucial point is that the cosmic rays isotropise on either side of the shock due to small-angle scattering off magnetic field fluctuations. The faster the isotropisation, the faster the particle can recross the shock. Every time the shock is crossed, a net energy gain is received by the particle crossing the shock. Although the acceleration efficiency depends on the effective scattering efficiency, the resulting spectrum is independent of the diffusion coefficient. In many cases the most efficient scattering rate of Bohm diffusion, where the mean free path is of the order of the gyroradius, is used in order to generate cosmic rays with the high energies that are observed.

An intuitive way of approaching the problem is by evaluating the number of particles that is located at the shock versus the number of particles that escape downstream. Only particles that do not escape qualify for the next round of acceleration.This is the approach originally described by \citet{1978Bella} and described as the microscopic approach in the review by \citet{1983Drury}. The macroscopic approach, originating from \citet{1977Krymskii, 1977Axfordetal, 1978BlandfordOstriker}, derives the acceleration and resulting powerlaw from the distribution function, requiring continuity at the shock. Below we will briefly summarise both methods and we refer to the original papers or \citet{1983Drury} for a more extensive treatment.

The distribution of relativistic particles can be described by the Vlasov-Fokker-Planck equation:
\begin{eqnarray}
\frac{\partial f}{\partial t} + (u+v)\frac{\partial f}{\partial z} - \frac{1}{3}\frac{\partial u}{\partial z} p \frac{\partial f}{\partial p} - \frac{\partial }{\partial z}\left( \kappa(z,p)\frac{\partial f}{\partial z}\right) = 0
\end{eqnarray}

The distribution can be separated into an isotropic part, and anisotropic parts to arbitrarily high order:
\begin{eqnarray}
f=f_0+\frac{f_i v_i}{v}+\frac{f_ij v_i v_j}{v^2} + \ldots
\end{eqnarray}
Mostly, the diffusion approximation is used, in which the first order anisotropy ($f_1=f_i v_i /v$) is used and eliminated, to arrive at a distribution that depends on the isotropic cosmic ray density ($f_0$) alone. 

\begin{eqnarray}
\frac{\partial f_0}{\partial t} + u\frac{\partial f_0}{\partial z} + \frac{c}{3}\frac{\partial f_1}{\partial z}- \frac{1}{3}\frac{\partial u}{\partial z} p \frac{\partial f_0}{\partial p} = 0\\
c \frac{\partial f_0}{\partial z}=-\nu f_1\\
\frac{\partial f_0}{\partial t} + u\frac{\partial f_0}{\partial z} -\frac{\partial}{\partial z}\left( \frac{c^2}{3 \nu}\frac{\partial f_0}{\partial z}\right)- \frac{1}{3}\frac{\partial u}{\partial z} p \frac{\partial f_0}{\partial p} = 0
\label{eq:diff_approx}
\end{eqnarray}
where we used $\kappa=c^2/(3 \nu)$.

From Eq.~\ref{eq:diff_approx}, far upstream the steady state solution ($\partial_t=0$ and $\partial u/\partial z=0$) implies that $f_0$ should have the form
\begin{eqnarray}
f=f_0(p) {\rm e}^{-\left(\frac{3 u_1 \nu}{c^2} \right) z}
\end{eqnarray}
to have a bound solution, where we use that in the shock frame $u=-u_1$. Or alternatively, 
\begin{eqnarray}
\frac{c}{3}f_1= u_1 f_0.
\end{eqnarray}
Downstream the steady state solution gives
\begin{eqnarray}
f=f_0(p).
\end{eqnarray}
At the shock the solutions have to connect, giving the boundary condition:
\begin{eqnarray}
\frac{c}{3} f_1 + \frac{u_1}{3}p\frac{\partial f_0}{\partial p}=\frac{u_2}{3}p\frac{\partial f_0}{\partial p}.
\end{eqnarray}
Using the boundary condition for the far upstream we can replace $c f_1/3$ with $u_1 f_0$, giving:
\begin{eqnarray}
u_1 f_0 +\left(\frac{u_1}{3}-\frac{u_2}{3}\right)p\frac{\partial f_0}{\partial p}=0,
\end{eqnarray}
which results in requiring that the cosmic ray density follows a powerlaw distribution:
\begin{eqnarray}
f_0 \propto p^{-q}
\end{eqnarray}
with
\begin{eqnarray}
q=3 u_1/(u_1-u_2) = 3r/(r-1),
\end{eqnarray}
where $r=u_1/u_2$ represents the compression ratio at the shock.
This powerlaw is valid in the test particle approach, for a planar shock, where the magnetic field is parallel to the shock normal. More details can be found in the original papers \citep{1977Krymskii, 1977Axfordetal, 1978BlandfordOstriker}. 

An alternative approach was used by \cite{1978Bella} to derive the powerlaw distribution of shock accelerated particles. It is based on the microscopic physics and is helpful to get insight in how the powerlaw may change depending on escape probability and probability of crossing the shock, which will be useful in understanding the physics of the later sections.

The flux of particles downstream is just the number of particles $n$ that are advected with the downstream flow velocity $u_2$: $n u_2$. The number of particles crossing the shock front per unit time from upstream to downstream in case of an isotropic distribution is half the number of particles moving towards the shock, and their average velocity over angle is again half of the shock velocity, giving for the flux $n c/4$. The fraction of particles not returning to the shock is therefore $n u_2/(nc /4)=4 u_2/c$. The probability of recrossing the shock can be high: $P_{ret}=1-4 u_2/c$. 

The energy gain of a particle crossing the shock from upstream to downstream can be calculated by transforming the momentum of the particle in the upstream to the downstream frame: $p'=p(1+(u_1-u_2) \cos \theta/c)$ such that the average change in momentum is $2 p(u_1-u_2)/(3 c)$. The energy gain from downstream to upstream is exactly the same, as $u_1$ and $u_2$ are interchanged and the angle of integration runs to the opposite side, yielding an extra $-1$. Thus the gain of momentum after a complete cycle is $\Delta p=4p(u_1-u_2)/(3c)$. After $k$ cycles, the number of particles has decreased as $n=n_0(1-4 u_2/c)^k$ and the momentum has increased as $p=p_0(1+4(u_1-u_2)/(3c))^k$. The number of particles as a function of momentum can be found to be 
\begin{eqnarray}
\frac{\ln(n/n_0)}{\ln(p/p_0)}=\frac{k \ln(1-4 u_2/c)}{k \ln(1+4(u_1-u_2)/(3c)}\approx \frac{-4u_2/c}{4(u_1-u_2)/(3c)}=-\frac{3}{r-1},
\end{eqnarray}
where $r$ again is the compression ratio $r=u_1/u_2$, such that:
\begin{eqnarray}
\frac{n}{n_0}=\left(\frac{p}{p_0}\right)^{-3/(r-1)}
\end{eqnarray}
and the differential energy spectrum is:
\begin{eqnarray}
n {\rm d}p \propto p^{-(r+2)/(r-1)} {\rm d}p.
\end{eqnarray}
In terms of the distribution function we arrive at the same answer as from the macroscopic approach, since:
\begin{eqnarray}
f_0=\frac{n}{4 \pi p^2}\propto p^{-3r/(r-1)}.
\end{eqnarray}

For a more detailed treatment we refer to the original papers and earlier reviews \citep{1978Bella, 1978Bellb, 1983Drury}.

\section{Magnetic field amplification: resonance regime}
\label{sec:mfa}

In the theory of diffusive shock acceleration cosmic rays are accelerated
by crossing and recrossing a shock, as shown in the previous section. 
On each cycle of crossing recrossing between upstream and downstream
the cosmic ray energy increases by a small fraction $\sim u_s/c$ where $u_s$ is the shock velocity.
For acceleration to PeV energies a cosmic ray (CR) has to cross the shock 
$\sim 10c/u_s$ times.
A shock propagating into a purely uniform magnetic field cannot accelerate
CR to PeV because charged particles pass easily through the shock and escape upstream or downstream
making only one pass through the shock and gain little energy.
Arguably, it was the realisation that charged particles are not free to escape the shock
environment
that provoked the development of the theory of shock acceleration in the late 1970's.

Fast and efficient CR acceleration by the Fermi mechanism
requires that particles are multiply scattered by magnetic
fluctuations in the acceleration source (e.g. shock). Magnetic field amplification due to the
resonant cosmic-ray streaming instability was studied in the context
of galactic cosmic-ray origin and propagation since the 1960s
\citep[see e.g.][]{1971KulsrudCesarsky,1974Wentzel,1981Achterberg,1990Berezinskiietal, 2003Zweibel}.
It was proposed by \citet{1978Bella} as a source of magnetic turbulence
in the test particle DSA scenario.

CR streaming along magnetic field
lines excite unstable growth of Alfven waves with wavelengths comparable with the CR Larmor radius \citep{1969KulsrudPearce, 1974Wentzel, 1975Skillinga, 1975Skilling, 1975Skillingc}. 
The Alfven waves consist of circularly polarised distortions to the magnetic field lines.
CR gyrating along the field lines in spatial resonance with the fluctuations are strongly scattered
and consequently execute a random walk along a field line.
The appropriate model for CR transport 
in the shock environment is diffusion instead of free propagation.
CR cross a shock many times with a statistical probability that 
naturally results in a $E^{-2}$ energy spectrum for cosmic rays \citep{1977Krymskii, 1977Axfordetal, 1978Bella, 1978BlandfordOstriker}, as shown in Section~\ref{sec:dsa}.

The theory of wave excitation and CR scattering had previously been
applied to CR propagation through the Galaxy.
Skilling set out the coupled equations for the wave energy density $I$ and the CR distribution function $f$.  When applied to a steady state wave and CR precursor
ahead of a shock the equations take the form:
\begin{eqnarray}
u_s\frac{\partial f}{\partial z}
=\frac{\partial }{\partial z}\left ( \frac{4}{3\pi} \frac{c r_g}{I} \frac{\partial f}{\partial z} \right )
\hskip 1.3 cm
u_s\frac{\partial I}{\partial z}=\frac{v_A}{B_0^2/8 \pi}\frac{\partial (4\pi cp^4f/3)}{\partial z},
\label{eq:resinst} 
\end{eqnarray}
where $v_A$ is the Alfven speed, $p$ is the CR momentum, and $r_g$ is the 
CR Larmor radius. 
In back-of-the-envelope terms, 
$4\pi cp^4f/3$ is the CR pressure. $I$ is the ratio of the energy density in the Alfven waves ($\delta B^2/4\pi$) to the energy density of the unperturbed field $B_0$, $I =2 \delta B^2/B_0^2$.
Further details of the equations can be found in \citet{1975Skillinga, 1975Skilling, 1975Skillingc},
and a solution of the precursor equations can be found in \citet{1978Bella}.
The dominant physics of the interaction between CR and the Alfven waves
is on the one hand that wave growth is driven by the CR pressure gradient 
and on the other hand that the CR diffusion coefficient
is inversely proportional to the wave energy density with mean free path $\Lambda$
given by $\Lambda = 4r_g/3I$. 
The equation for wave evolution can be integrated to give 
\begin{eqnarray}
I=2 \frac{u_s}{v_A} \frac {4\pi c p^4 f/3}{\rho u_s^2}
\label{eq:intensity} 
\end{eqnarray}
For a characteristic interstellar magnetic field ($B \sim 3 \mu$G) 
the Alfven speed is around $v_A \sim 10$~km s$^{-1}$, 
so $u_s/v_A$ is of the order of $10^{3}$ for the
outer shock of a young supernova remnant (SNR).
To account for the energy density of Galactic CR,
acceleration by SNR must be efficient and the energy transfer into
CR at a SNR shock has to be in the range $0.1-0.5\rho u_s^2$ 
\citep{1934BaadeZwicky,2010Longair}.
In terms of Equation~\ref{eq:intensity} this gives
$I\gg 1$, which would correspond to a perturbed field $\delta B$ greatly exceeding 
the zeroth order field $B_0$.
Naively this implies a diffusion coefficient much less than $r_g c$ and a CR
scattering mean free path much less than the Larmor radius ($\Lambda \ll r_g$).
The linear instability depends upon a resonance between the CR Larmor radius and the Alfven wavelength. This
resonance is destroyed when $\delta B$ approaches $B_0$,
in which case the instability cannot be resonantly driven and the instability 
is expected to saturate at about $\delta B \sim B_0$.
The linear equations lose validity when $I\sim 1$, but Equation~\ref{eq:intensity}
suggests that instabilities driven by CR streaming
may be able to amplify magnetic field far
beyond its initial ambient value. In the next sections we will turn to instabilities that do not rely on the resonance condition and may continue well beyond $\delta B/B_0 \sim1$.

\section{Magnetic field amplification: short wavelength regime}
\label{sec:shortscale}
\subsection{A non-linear estimate of the amplified magnetic field}

A more basic understanding of the opportunity for magnetic field amplification
can be derived as follows.
According to equation \ref{eq:resinst} 
the rate at which the wave energy density $I$ grows is 
$\approx v_A \partial P_{cr}/\partial z$ where $P_{cr}$ is the CR pressure.
This can be interpreted as a force $ \partial P_{cr}/\partial z$ pushing against 
magnetic fluctuations at the Alfv\'en velocity $v_A$.
Provided the CRs are coupled to the magnetic field fluctuations and the fluctuations
propagate at about the Alfv\'en speed this equation is 
approximately valid even when the spatial
resonance between the wavelength and the Larmor radius breaks down.

In the linear regime it makes sense to think of the magnetic fluctuations moving at the Alfv\'en velocity.  
In the non-linear regime, the fluctuations no longer take the form of linear waves, 
but still they can be estimated to move 
at a velocity $v_f \sim v_A \sim({\rm magnetic\ pressure / density})^{1/2}$.
Energy transfer to the fluctuations occurs at a rate 
$v_f \partial P_{cr}/\partial z$,
and the equation for growth of the turbulent energy density $U_f$ associated
with the fluctuations is
\begin{eqnarray}
u_s \partial U_f/\partial z
=v_f \partial P_{s}/\partial z,
\label{eq:dudz}
\end{eqnarray}
where $P_s$ is the CR pressure at the shock.
Assuming that $v_f=(U_f/\rho)^{1/2}$ and $U_f=B^2/4 \pi$ as for Alfv\'en waves, the turbulent energy density at the shock is
\begin{eqnarray}
\frac {B^2}{4 \pi} \approx \frac{1}{4} \left ( \frac{P_{cr}}{\rho u_s ^2} \right )^2 \rho u_s^2
\label{eq:bpropu}
\end{eqnarray}

If the CR acceleration efficiency is $P_{s}\approx 0.1 \rho u_s^2$ this
leads to an estimated magnetic field of the order of $100\ \mu$G in young SNR
in good agreement with observations \citep{2003VinkLaming, 2003Berezhkoetal, 2005Voelketal}.
For a fixed acceleration efficiency $P_{cr}\propto \rho u_s^2$,
the amplified magnetic field is proportional to $\rho^{1/2} u_s$, see \citet{2001BellLucek} for a more detailed model.
A later improved analysis of field amplification
suggests a stronger dependence
on shock velocity, $B \propto \rho^{1/2} u_s^{3/2}$, 
as discussed in Sect.~\ref{sec:nonlinbell} \citep{2004Bell, 2009Bell}. 

\subsection{A non-resonant instability}
\label{sec:nonresonantbell}

Equation \ref{eq:bpropu} sidesteps
the requirement for a resonance between the Larmor radius and the instability wavelength
and subsequent theoretical developments have shown that the resonance is not essential
for the amplification of magnetic field by CR streaming.
The same linearisation of the Vlasov equation that gives rise to the resonant instability also 
identifies a non-resonant instability 
as first demonstrated by \citet{2004Bell}.
\citet{2004Bell} simplified the problem by using the MHD equations to treat the thermal plasma
as a magnetised fluid.
A kinetic Vlasov treatment is retained for the CR.
The CR exert a force on the MHD plasma through the reaction on the 
${\bf j}_{cr}\times {\bf B}$ force, which is equal and opposite in most of the unstable regime, the details of which will be discussed in Sect.~\ref{sec:returncurrent}. Here $j_{cr}$ is the CR electric current
density, and the CR trajectories are calculated in the magnetic and electric fields
calculated by the MHD model.

The coupled Vlasov-MHD equations give a linear dispersion relation containing two
different instabilities for the two 
different circular polarisations and wavenumbers ${\bf k}$ parallel
to the zeroth order magnetic field.
The resonant instability instability occurs when the circular polarisation is such
that CR moving in the same direction as $j_{cr}$ gyrate resonantly around the zeroth
order field in the same sense as the helical perturbations in the magnetic
field, which we will refer to as the left-hand polarisation. 
A stronger non-resonant instability is found in the opposite (right-hand) circular polarisation.

\begin{figure}
\begin{center}
\includegraphics[width=\textwidth]{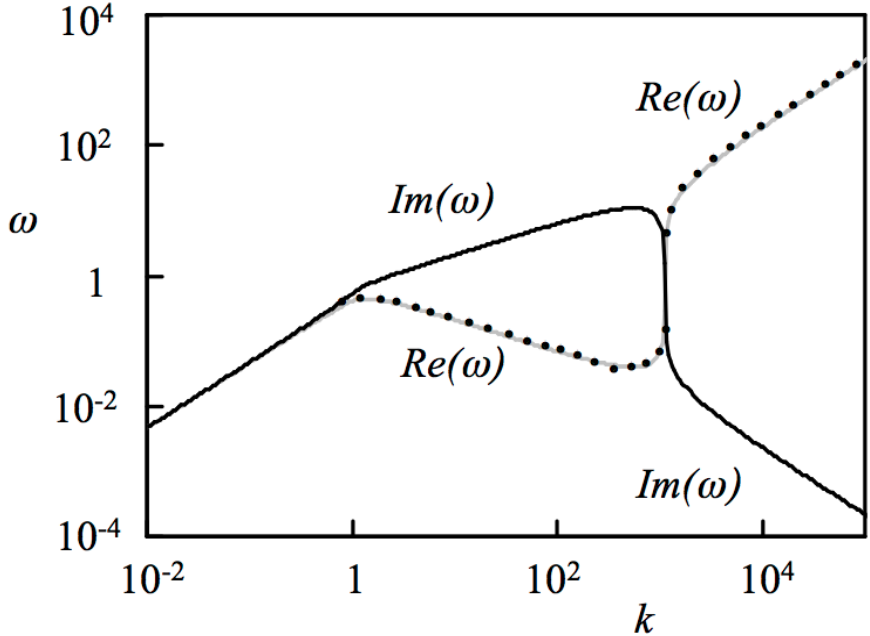}
\caption{ Dispersion relation for the non-resonant instability
$v_A=6.6\times 10^3 {\rm ms}^{-1}$, $u_s/c=1/30$.
$\omega $ is in units of $u_s^2/cr_{g}$ and $k$ in units of $r_{g}^{-1}$
where $r_g$ is the Larmor radius of the lowest energy CR.
Reproduced from \citet{2004Bell}.
}
\label{fig:bellinstability}
\end{center}
\end{figure}

The dispersion relation for the non-resonant  instability
is plotted in Figure~\ref{fig:bellinstability} for a power-law CR distribution $f(p)\propto p^{-4}$
for $p_1<p<p_2$ where $f(p)$ is non-zero between momenta $p_1$ and $p_2$.
The resonant and non-resonant instabilities have similar growth rates when $k \approx r_g^{-1}$,
that is when the inverse wavenumber $k^{-1}$ is 
approximately equal to the Larmor radius of CR with the lowest momentum $p_1$.
The lowest momentum CR are the most important because they carry most of the electric current.
At wavelengths less than $r_g$ there are few CR in spatial resonance with
the helices in the magnetic field and the growth rate of the resonant
instability decreases as $k$ increases.
In contrast, the non-resonant instability does not depend on resonance and its
growth rate instead increases with increasing $k$.
The non-resonant growth rate increases with $k$ until tension in the magnetic
field overpowers the driving force, that is when 
 ${\bf j}_{cr}\times {\bf B}={\bf B}\times ({\bf k}\times {\bf B})c/4\pi$
 and $kB \sim 4\pi j/c$.

 The essential difference between the resonant and non-resonant linear instabilities
 can be explained as follows.
 In both cases the background MHD fluid motions
 are driven by the  ${\bf j}_{cr}\times {\bf B}$ force.
 ${\bf j}_{cr}$ and ${\bf B}$ have unperturbed zeroth order components,
  ${\bf j}_{cr0}$ and ${\bf B}_0$, 
  and first order perturbed components
  ${\bf j}_{cr1}$ and ${\bf B}_1$ respectively.
  The ${\bf j}_{cr}\times {\bf B}$ driving force
  has two first order components, ${\bf j}_{cr0}\times {\bf B}_1$
 and ${\bf j}_{cr1}\times {\bf B}_0$.
 The resonant instability is driven by ${\bf j}_{cr1}\times {\bf B}_0$ where
 ${\bf j}_{cr1}$ is the perturbed CR current which is especially strong when
 the CR trajectories respond resonantly to a helical magnetic field with a wavelength
 equal to the CR Larmor radius.
 On the other hand, the non-resonant instability is driven by 
 the other first order force
 ${\bf j}_{cr0}\times {\bf B}_1$.  
 In the non-resonant instability 
 only the uniform zeroth order current ${\bf j}_{cr0}$ matters
and there is no requirement for a resonance with the CR Larmor radius.

Because the non-resonant instability is driven by the zeroth order CR current,
we can derive a growth rate while omitting
the first order current from the analysis.
The response of the CR trajectories to the perturbed fields
can be ignored, and hence there is no need to solve the Vlasov equation
for CR.
It is sufficient to solve the MHD equations for the thermal plasma
with sole addition of the 
${\bf j}_{cr0}\times {\bf B}_1$ force in the MHD momentum equation.
The MHD mass conservation equation can also be omitted since the instability
is transverse and the density is unperturbed, $\rho _1=0$.
Similarly the thermal plasma pressure is unperturbed.
The first order equations are then
\begin{eqnarray}
\rho \frac {\partial {\bf u}_1}{\partial t}=\frac{-{\bf j}_{cr0}\times {\bf B}_1}{c}-
\frac{1}{4\pi} {\bf B}_0\times (\nabla \times {\bf B}_1)
\hskip 1.5 cm
\frac{\partial {\bf B}_1}{\partial t}=
\nabla \times ({\bf u}_1 \times {\bf
B}_0)
\label{eq:mhd1bell} 
\end{eqnarray}
${\bf u}_1$ can be eliminated between the equations to give
\begin{eqnarray}
\frac{\partial {\bf B}_1^2}{\partial t^2}
-\frac{1}{4\pi \rho} \left ({\bf B}_0 \cdot \nabla \right )^2 {\bf B}_1
=
-\frac{1}{\rho c}
\left ({\bf B}_0\cdot \nabla \right )\left ({\bf j}_{cr0}\times {\bf B}_1 \right
)
\label{eq:dbdt2} 
\end{eqnarray}

Harmonic solutions 
${\bf B}_1 \propto \exp[i(kz-\omega t)]$ with 
${\bf k}$ parallel to ${\bf B}_0$ in the $z$ direction, 
gives a dispersion relation
\begin{eqnarray}
\omega ^2-k^2 v_A^2= \pm  \frac{kB_0 j_{cr0}}{\rho c}
\label{eq:dispnr} 
\end{eqnarray}
In the absence of a CR driving current $j_{cr0}=0$, this
is the dispersion relation for an Alfv\'en wave.
The right hand side of the dispersion relation creates an instability
provided ${kB_0 j_{cr0}}/{\rho c}>k^2 v_A^2$ and the negative sign is chosen for $\pm$
corresponding to the non-resonant polarisation.
If $k^2 v_A^2 > |  {kB_0 j_{cr0}}/{\rho c}|$, the tension in the field lines
overpowers the driving term and the dispersion relation is that
for non-growing Alfv\'en waves with a modified phase velocity.
The neglect from the analysis of the 
${\bf j}_{cr1} \times{\bf B}_0$ 
ignores the tendency of CR to follow the field lines on wavelengths greater than the CR Larmor radius
and therefore omits the reduced instability for $kr_{g}<1$ that is seen in
Figure~\ref{fig:bellinstability}.
For wavenumbers $k$ in the unstable range 
$r_g^{-1}<k<4 \pi j_{cr0}/B_0 c$, equivalent to 
$r_g^{-1}<k<B_0j_{cr0}/(\rho v_A^2 c)$, the instability is purely growing
with a growth rate
\begin{eqnarray}
\gamma = \left (  \frac{kB_0 j_{cr0}}{\rho c}    \right )^{1/2}
\label{eq:gammanr}
\end{eqnarray}
and a maximum growth rate, determined by the tension in the magnetic field,
\begin{eqnarray}
\gamma _{max}=\frac{1}{2}\left( \frac{4\pi}{\rho} \right )^{1/2}\frac{j_{cr0}}{c}
\label{eq:gammamaxnr} 
\end{eqnarray}
The CR current $j_{cr0}$ in the precursor is related to the CR pressure
by 
$$j_{cr0}=e(u_s/c)P_{cr}/p_1\ln(p_2/p_1).$$
If the CR acceleration efficiency is defined by
$P_{cr}=\eta \rho u_s^2$, the number of e-foldings for the fastest growing mode is
\begin{eqnarray}
\gamma _{\rm max} \tau_{\rm pc}= \frac{\kappa}{\kappa_B} \frac{\eta M_A} {2 \ln(p_2/p_1)}
\label{eq:growthnr} 
\end{eqnarray}
where $\omega _g$ is the CR Larmor frequency, $M_A$ is the Alfven Mach
number $M_A^2=4\pi \rho u_s^2/B^2$, 
$\kappa/\kappa_B$ is the ratio of the CR diffusion coefficient to the
Bohm coefficient,
and $\tau_{\rm pc}=({c^2}/{u_s^2}) \omega _g^{-1}$ is the time taken for the
shock
to propagate a distance equal to the scaleheight of the CR precursor if Bohm
diffusion applies.
Since $M_A \sim 10^3$ for a young SNR propagating into an interstellar
magnetic field, and overall CR efficiencies must be $10-50$\%,
the number of instability e-folding times is much greater than one.
The term $\kappa/\kappa_B$ introduces an element of self-regulation since
the number of e-foldings is greater if the magnetic fluctuations are small and
diffusion is greater than Bohm. 
This may be important at the foot of the precursor where the 
instability has not had the opportunity to grow.
If scattering is weak, CR escape a relatively large distance upstream,
initiate instable growth far ahead of the shock and remedy the
lack of a perturbed magnetic field able to scatter the CR.

\subsection{Return currents and energy transfer}
\label{sec:returncurrent}

In the above analysis the force on the background thermal plasma was included
as $-{\bf j}_{cr}\times {\bf B}$.
To conserve momentum 
the force on the background plasma is equal and opposite to the force
on the CR.  
Another way of looking at this is that the force
on the background plasma is exerted through the return current ${\bf j}_t$
that the background thermal
plasma must carry to neutralise the current carried by the CR, 
${\bf j}_t\approx -{\bf j}_{cr} $.
The difference between ${\bf j}_t$ and $-{\bf j}_{cr} $
is given by the Maxwell equation when the displacement current is neglected:
${\bf j}_t=-{\bf j}_{cr}+c \nabla \times {\bf B}/4 \pi$ 
in which case
${\bf j}_t\times {\bf B} =  -{\bf j}_{cr}\times {\bf B}-{\bf B}\times (\nabla
\times {\bf B})c/4 \pi$.
${\bf j}_t$ and ${\bf j}_{cr} $ very nearly cancel out where the instability is strong.
However, their non-cancellation at small wavelengths gives rise
to the magnetic tension that limits the instability to the range
$k<B_0j_{cr0}/\rho v_A^2 c$.
At longer wavelengths the thermal return current follows a path very close to that
of the CR current.

A necessary condition for CR-driven instability
is that the thermal particles are strongly magnetised with Larmor radii much smaller 
than the wavelength and that the CR should have a Larmor radius comparable with or larger
than the wavelength so they are not tied to magnetic field lines.
If the thermal particles are frozen-in to the magnetic field
it is not immediately obvious how one can have a ${\bf j}_t\times {\bf B}$
force with a current ${\bf j}_t$ unaligned with the magnetic field.
The difficulty is resolved through the theory
of cross-field drifts.
The ${\bf j}_t\times {\bf B}$ force imparts an acceleration to the plasma. This acceleration can be viewed as being equivalent to a gravitaional force. A charged particle in a gravitational field executes a cross field drift, and similarly in this case a cross field drift arises through the Lorentz force, which produces the current density 
${\bf j}_t$.

Another conundrum is how energy is extracted from the
CR current to drive the turbulence and amplify the magnetic field.
The ${\bf j}_{cr}\times {\bf B}$ cannot extract energy from the CR because
magnetic field only deflects particles and cannot change their energy.
The solution here comes from the second order electric field 
$c {\bf E}_2=-{\bf u}_1 \times {\bf B}_1$.
${\bf E}_2$ is anti-parallel to ${\bf j}_{cr0}$ and
reduces the CR energy which is transferred to the second order
magnetic energy density ${\bf B}_1^2 /8 \pi$ and kinetic energy density
$\rho {\bf u}_1^2 /2$.

\subsection{ The non-resonant instability in 3 dimensions}

In the case of a parallel shock, the zeroth order CR current is 
parallel to the zeroth order magnetic field and the above theory
for ${\bf k}$ aligned with ${\bf B}_0$ can be applied directly.
A more general theory is needed for an
oblique shock where ${\bf k}$ and ${\bf B}_0$ are not parallel.
The dispersion relation for general orientations of
${\bf k}$, ${\bf j}$ and ${\bf B}_0$ was derived by \citet{2005Bell} for
wavelengths shorter than the CR Larmor radius, ie
$kr_g>1$:
\begin{eqnarray}
&&\left\{ \gamma ^2 +(\hat {\bf k}\cdot\hat {\bf b})^2 k^2 v_A^2\right \}
\left\{ \gamma ^4 +\gamma^2 k^2 (v_A^2+c_s^2)+
(\hat {\bf k}\cdot\hat {\bf b})^2  k^4v_A^2 c_s^2\right \}
\\\nonumber
&&=\gamma_0^4\left \{ \gamma^2+ (\hat {\bf j}\cdot\hat {\bf k})^2 k^2 c_s^2
+\left [ (\hat {\bf k}\cdot\hat {\bf b})^2 + (\hat {\bf k}\cdot\hat {\bf j})^2
-2(
\hat {\bf k}\cdot\hat {\bf j}) (\hat {\bf b}\cdot\hat {\bf j}) (\hat {\bf k}\cdot\hat {\bf b}) \right ]
k^2 v_A^2 \right \}
\end{eqnarray}
where $v_A=B_0/(4\pi \rho_0)^{1/2}$ is the Alfven speed,
$c_s=(\partial P/\partial \rho)^{1/2}$ is the sound speed,
$\gamma_0^4=( {\bf k}\cdot {\bf B}_0)^2 {\bf j}^2/\rho_0^2 c^2$,
and a hat denotes a unit vector: $\hat{\bf k}={\bf k}/|{\bf k}|$,
 $\hat{\bf b}={\bf B}_0/|{\bf B}_0|$,
 $\hat{\bf j}={\bf j}_{cr}/|{\bf j}_{cr}|$.
 The terms involving $kv_A$ are important at the short wavelength limit 
 when magnetic field tension is important as in the case of ${\bf k}$ aligned with ${\bf B}_0$.
 The terms involving $kc_s$ represent additional short wavelength compressibility effects that are not
 present when $\bf k$, ${\bf B}_0$ and ${\bf j}_{cr}$ are all parallel.
The terms in $kv_A$ and  $kc_s$ are important only at short wavelengths.
At longer wavelengths they can be neglected and the growth rate simplifies
to
\begin{eqnarray}
 \gamma = \left [ \frac {({\bf k}\cdot{\bf B}_0)  j_{cr}}{\rho_0 c} 
 \right ]^{1/2}
 \label{eq:bellgamma}
\end{eqnarray}
 The instability grows most rapidly for wavenumbers parallel to the magnetic
field but the growth rate is independent of the mutual orientation of the magnetic field ${\bf B}_0$ and the CR current ${\bf j}_{cr}$.

The insensitivity of the growth rate to the angle between magnetic field
and CR current implies that the instability is present for perpendicular as well as parallel shocks.
In fact, the growth rate is faster for perpendicular shocks 
because the CR current is larger than that upstream of parallel shocks.
In the precursor of a parallel shock the CR drift at the shock velocity
relative to the thermal plasma to give a current density parallel to the
shock normal of $j_{cr}=n_{cr}eu_s$ where $n_{cr}$ is the CR number
density. In contrast, for a perpendicular shock the CR current density is larger in the  direction perpendicular to the shock normal.
The CR current density at a perpendicular shock can be calculated from the
first moment of the Vlasov-Fokker-Planck (VFP) equation in the diffusive limit in which
the one-dimension CR distribution function takes the form
$f({\bf p},z,t)=
f_0(|{\bf p}|,z,t)+{\bf f}_1(|{\bf p}|,z,t)\cdot ({\bf p}/|{\bf p}|)$.
\begin{eqnarray}
\frac{\partial {\bf f}_1}{\partial t}
+c\frac{\partial {\bf f}_0}{\partial z}
+\frac{ec}{p}{\bf E} \frac{\partial f_0 }{\partial p}
+\frac{e{\bf B}}{p} \times {\bf f}_1=-\nu {\bf f}_1
\end{eqnarray}
where $\nu$ represents angular scattering by small scale fluctuations.
For a mono-energetic CR distribution at 
a perpendicular shock, the CR current density in the precursor
can be separated into a component
$j_{||}$ normal to the shock 
and a component $j_\perp$ that is perpendicular to the
magnetic field and the shock normal:
\begin{eqnarray}
j_{||}= \frac{1}{3} \frac{\nu \omega _g}{\omega _g^2 +\nu ^2}\frac{r_g}{L} n_{cr} e c
\hskip 1 cm
j_\perp= \frac{1}{3} \frac{\nu \omega _g}{\omega _g^2 +\nu ^2}\frac{\Lambda }{L} n_{cr} e c
\end{eqnarray}
where $n_{cr}$ is the CR number density, $L$ is the scalelength of the precursor, $\Lambda=c/\nu$ is the CR mean
free path, and $r_g=c/\omega _g$ is the CR Larmor radius.

The components $j_{||}$ and $j_\perp$ of the CR current density
correspond to the CR drift velocities in the local upstream fluid
rest frame.
In steady state, $j_{||}=n_{cr}e u_s$  and $j_\perp=(\Lambda /r_g) n_{cr}e u_s$.
The large scale field is relatively unimportant
when the mean free path is comparable with the Larmor radius 
$\Lambda \sim r_g$ (Bohm diffusion) since
the distinction between parallel and perpendicular shocks is
then relatively minor, and the scaleheight in both cases is $L\sim(c/u_s) r_g$.
Bohm diffusion corresponds to the smallest possible mean free path.
More usually, the mean free path can be expected to be greater than
the Larmor radius and the cases of parallel and perpendicular shocks become quite different.
From the above equations, the precursor scaleheight ahead of a perpendicular shock is reduced to
$ L\sim (r_g/\Lambda)(c/u_s) r_g$,
and $j_\perp$ exceeds $j_{||}$.
$j_\perp$ 
results from the non-cancellation of gyratory currents 
in the CR density gradient in the precursor.
Since $j_\perp$ is greater
than $j_{||}$, the instability is driven more rapidly at a perpendicular shock
than at a parallel shock.
However, the time during which
the instability can grow is reduced because the scaleheight $L$ is 
smaller at a perpendicular shock.
The increased growth rate and the reduced time for growth cancel out
and the number of linear e-foldings is the same for both
parallel and perpendicular shocks.  
Hence the non-resonant instability is equally effective for all
shocks whether they are perpendicular, parallel or oblique.
From Equation~\ref{eq:bellgamma}, in any of these cases fastest linear growth occurs for wavenumbers
parallel to the large scale magnetic field, independent of its orientation to the shock normal.
Unstable growth and magnetic field amplification
at perpendicular shocks has been demonstrated numerically in 
particle-in-cell simulations by \citet{2010RiquelmeSpitkovsky}.

\subsection{Non-linear magnetic field amplification}
\label{sec:nonlinbell}

A linear instability takes a small perturbation $\delta {\bf B}$ on the magnetic field
and amplifies it until it becomes comparable with the zeroth order field ${\bf B}_0$. 
When $\delta {\bf B} \sim {\bf B}_0$, the linear assumption that 
second order terms in $\delta {\bf B}$ can be neglected becomes untenable.
A crucial question is whether the non-linear terms cause the
instability to saturate and stop growing,
or whether the magnetic field grows further to a magnitude 
much greater than ${\bf B}_0$.
In the context of CR acceleration, the field has to 
continue growing beyond  $\delta B/B_0 \sim 1$
if diffusive shock acceleration is to explain the presence of PeV CR
in the Galaxy.
Furthermore, saturation at $\delta B/B_0 \sim 1$ is insufficient to explain the
large magnetic fields inferred from X-ray observations of synchrotron
emission at an SNR shock.
Therefore it is crucial to ascertain not only that linear growth
is sufficiently fast but also that the instability continues to grow
non-linearly beyond $\delta B/B_0 \sim 1$ to generate fields exceeding
$100\ \mu$G at SNR shocks.

Fortunately the non-resonant instability has the unusual 
property of continuing rapid growth into the non-linear regime.
Remarkably, in the restricted geometry of a monochromatic circularly polarised wave
with wavenumber ${\bf k}$, zeroth order uniform magnetic field  ${\bf B}_0$
and uniform CR current ${\bf j}_{cr}$
all parallel, the linear equations remain valid 
into the non-linear regime and the
instability continues to grow exponentially to arbitrary amplitude 
at the linear growth rate.
In this special case,
slabs of plasma with frozen in magnetic field 
continue to be accelerated in directions perpendicular to ${\bf k}$,
${\bf B}_0$ and ${\bf j}_{cr}$. 
In practice other modes with different ${\bf k}$ also grow
and these interfere to slow the growth.
For example, an exponentially expanding spiral field in one part
of the plasma is likely to collide with an expanding spiral field
seeded in a different part of the plasma.
The spirals cannot in general pass through each other and their
growth is limited.
However, the presence of exponentially growing non-linear modes
is a strong hint that growth to large amplitude is possible.
\citet{2000LucekBell} showed numerically that the magnitude of the magnetic field
can increase by at least an order of magnitude.
3D MHD simulations by \citet{2004Bell} showed similar or larger growth
before the calculation was terminated when magnetic structures
expanded to the size of the periodic computational box.
In \citet{2004Bell} the instability grew exponentially at
the expected rate until $\delta B/B_0 \sim 1$, whereafter it continued to grow
but more slowly.

\begin{figure}
\begin{center}
\includegraphics[width=\textwidth]{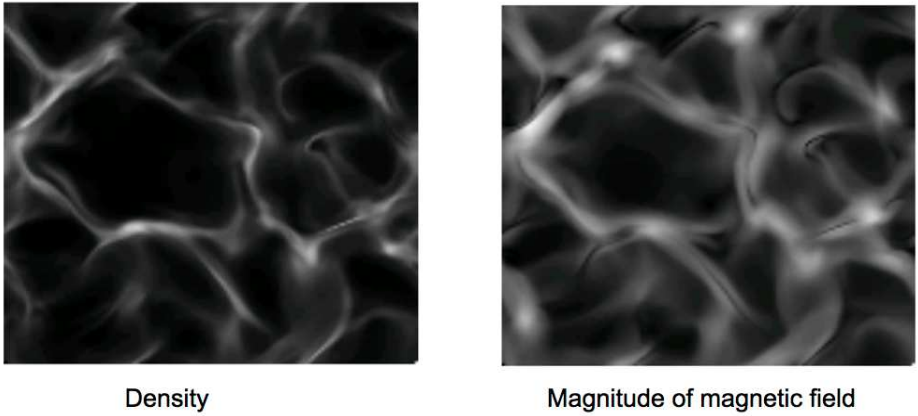}
\caption{ Comparison of structures of density and magnitude of magnetic field.
2D slice through a 3D simulation. Reproduced from \citet{2005Bell}.
}
\label{fig:nonlinearbell}
\end{center}
\end{figure}

The linear eigenmodes of the instability consist of
spirals of magnetic with a preferred helicity.
The evolution of this basic configuration into the non-linear regime can be seen in figure 4 of \citet{2004Bell}. 
Initially small spirals and loops of magnetic field grow non-linearly in radius.
Collisions between neighbouring spirals produce walls
of strong magnetic field surrounding cavities of very weak magnetic field.
The field is far from uniform and does not conform to
conventional pictures of randomly phased Fourier modes in $k$-space.
Because the magnetic field is frozen in to the background plasma, 
the density and magnetic field have closely correlated structures 
of walls and cavities as shown
in Figure~\ref{fig:nonlinearbell}, which reproduces two frames from figures 2 and 3 of \citet{2005Bell}.
The same wall-cavity struture has been found by \citet{2008Revilleetal,2008Zirakashvilietal} in MHD simulations,
and by \citet{2009RiquelmeSpitkovsky} and \citet{2009Ohiraetal}
in particle-in-cell simulations.

The growth of large scale structures resulting from the expansion of 
the small scale structures provides a natural way
of producing structures on the scale of a CR Larmor radius.
These are especially important since CR are most effectively scattered
by fields on this scale.
Fields on smaller scales may explain the amplified fields observed at SNR shocks,
but they cannot by themselves provide the strong CR scattering needed to accelerate
CR to PeV energies.
\citet{2008Revilleetal} have modelled CR transport in non-linear CR-driven 
magnetic fields calculated with a 3D MHD code.  
They show that the amplified field does inhibit CR transport and reduces diffusion
to less than Bohm diffusion in the initial magnetic field.
The generation of magnetic structures on large scales is an active field of research
(see section \ref{sec:longwavelength}), and other sources of turbulence contribute to the overall
magnetic field structure in SNR \citep{2007GiacaloneJokipii, 2008ZirakashviliPtuskin, 2009Beresnyaketal, 2009Inoueetal,2009Schureetal}.

It is clear from analyses and simulations that magnetic field amplification
continues far into the non-linear regime.
In some circumstances amplification may be limited by the time
for which the ${\bf j}_{cr}\times {\bf B}$ driving force operates.
In the case of a shock precursor this limiting time is the time
$L/u_s$ it takes for the precursor of scalelength $L$ to be overtaken by the shock.
However, two conditions have emerged as requirements for unstable growth.
One is that the scale size $k^{-1}$ of magnetic structures 
should not exceed the CR Larmor radius, otherwise the CR follow
the field lines and ${\bf j}_{cr}\times {\bf B}$ becomes small
since ${\bf j}_{cr}$ is parallel to ${\bf B}$.
The second is that the magnetic tension 
${\bf B} \times (\nabla \times {\bf B})c/4 \pi \sim kB^2 c/4 \pi$
should not exceed the ${\bf j}_{cr}\times {\bf B}$
driving force.
These conditions reduce to
$k>eB/pc$ and $k<4 \pi j_{cr}/Bc$
respectively.
For both conditions to be satisfied simultaneously
we need
$B^2/4 \pi < p j_{cr}/e$.
Using the expression for $j_{cr}$ derived in the discussion
between equations \ref{eq:gammamaxnr} 
and \ref{eq:growthnr} 
in section \ref{sec:nonresonantbell}, this reduces to
an estimate for the saturated magnetic energy density producable 
in the non-linear phase of the non-resonant instability.
\begin{eqnarray}
\left (
\frac{B^2}{4\pi} \right )_{sat}
\sim \frac{u_s}{c}\frac{\eta \rho u_s^2}{\ln (p_2/p_1)}
\end{eqnarray}
This estimate is a good match to observations of the magnetic field in SNR \citep{2008Vinkb, 2009Bell}, 
but other explanations are possible for the same data \citep{2011Malkovetal}.

\section{Magnetic field amplification: long length scales}
\label{sec:longwavelength}

As shown in the previous section, the non-resonant Bell instability can act efficiently to amplify magnetic fields on scales smaller than the gyroradius. However, in order to accelerate cosmic rays to higher energies, the magnetic field should also be amplified on scales beyond the cosmic ray gyroradius, which is the `long-wavelength regime' discussed in this section. In 2011 various papers have been published on possible instabilities that act to amplify the magnetic field on these scales \citep{2011Bykovetal, 2011SchureBell, 2011RevilleBell}, on which we will focus in this section. 
We will also briefly discuss other long-wavelength instabilities (firehose and acoustic), but for a more extensive review refer to the original papers or reviews \citep[see e.g. ][]{1987BlandfordEichler, 2001MalkovDrury, 2011Bykovetalreview}.

\subsection{Current-driven stress-tensor instability}
\label{sec:stresstensor}

The small-scale instability is essentially a fluid instability. The cosmic ray current is regarded to be of such scales that it is unperturbed by the non-resonant growth of small-scale magnetic fields. When looking at growth of magnetic field on scales larger than the gyro-radius, this assumption no longer holds since the CR trajectories follow the field lines and perturbations to the current in perpendicular directions should be taken into account.

One way to determine this effect on the stability of the plasma is by including higher order anisotropic terms in the kinetic equation. The feedback between the cosmic ray particles and the magnetic field in the plasma proceeds through forces acting on the momentum equation. The ${\bf j} \times \Bv$ forces arise when components of the cosmic ray current, or rather the induced return current in the plasma, are perpendicular to local components of the magnetic field. Gradients in the perpendicular current require taking into account higher-order components of the distribution function, at least the stress tensor, effectively representing gradients in the current.

The distribution function of relativistic particles can be written as:

\begin{eqnarray}
\label{eq:kinetic}
\partial_t \fv + \vv \cdot \nabla \fv + \frac{e}{m}\left(\Ev + \frac{\vv \times \Bv}{c}\right)\cdot\nabla_v\fv = 
\nabla_v \cdot ({\bf D} \cdot \nabla_v \fv),
\end{eqnarray}
where $\fv$ is the particle distribution in phase space, $\vv$ the particle velocity, and ${\bf D}= \frac{v^2\nu}{2}\left({\bf I} - {\bf \hat n}{\bf \hat n}\right)$ the diffusion tensor with $\nu$ the collision frequency, ${\bf I}$ the identity matrix, and ${\bf \hat n}$ the unit vector in the direction of the corresponding tensor component. 

The collisions, represented in the parameter $\nu$, are not actual collisions in such tenuous systems, but effectively act in the same way. What we mean with collisions is the cumulative effect of small-angle scattering as a result of the Lorentz force of the perturbed current and magnetic field. When they are fluctuating on the same scale, the Lorentz force is effectively deviating the path of the cosmic rays. Multiple of these small-angle scatterings result in isotropisation. The length and time scales on which this occurs is represented in the parameter $\nu$, which represents the scattering frequency in terms of the particle velocity over the mean free path. In Bohm diffusion this is taken to be of the order of the gyrofrequency. The short-wavelength non-resonant instability acts to amplify the field on scales that can efficiently deflect the low-energy cosmic rays. In effect, the parameter $\nu$ thus describes the momentum exchange of the small-scale instability and can be used to determine its influence on the long range.

The distribution of cosmic rays is dominated by the isotropic component: $f_0$, but also contains an anisotropic part, to increasing order $\fv=f_0+\fv_1 \cdot \vv/v+\fv_2 \cdot \vv \vv/v^2 + \ldots$. $\fv_1$ can be viewed as the directional component of the cosmic ray distribution, or as the gradient of $f_0$, and acts like a current. $\fv_2$ is the pressure tensor, of which the isotropic part of the diagonal is normally included in the $f_0$ term. Anisotropy in the diagonal can be responsible for the firehose instability. Off-diagonal terms embody the stress-tensor and reflect gradients in the current. Each higher order is a factor of $\uv/c$ smaller than the previous order, where $\uv$ is the drift velocity.
Evaluation of the transport equation to zeroth order, being the isotropic part, and first and second order anisotropies, gives the following system of equations, where we ignore any contribution of higher (3rd) order:
\begin{eqnarray}
\label{eq:f0}
\partial_t f_0=-\frac{c}{3}\nabla \cdot \fv_1.\\
\label{eq:f1v}
\partial_t \fv_1 + c\nabla f_0+\frac{e}{mc}(\Bv \times \fv_1)+\nu \fv_1 +\frac{2}{5}c\nabla \cdot \fv_2 = 0.\\
\label{eq:f2v}
\left[c\left(\nabla \fv_1 - \frac{1}{3}\nabla \cdot \fv_1 {\bf I}_2 \right) + \frac{2e}{mc}(\Bv \times \fv_2) + 3 \nu \fv_2\right]_2=0.
\end{eqnarray}
Here ${\bf I}_2$ is the second order unity tensor, and $[\ldots]_2$ indicates a summation of the permutations for ${\bf ijk}$ in two ways divided by 2, such that we get a symmetric tensor with components that satisfy $f_{ij}=f_{ji}$. In principle also higher order terms can be used in the evaluation, but it turns out these have no significant effect on the instability \citep{2011SchureBell}.
This system of equations can be closed in combination with the MHD equations:
\begin{eqnarray}
\label{eq:maxwell}
\partial_t \Bv &=& \nabla \times (\uv \times \Bv)\\
\rho \partial_t  \uv&=&\frac{{\bf j}_{th} \times \Bv}{c}-\nabla P - \nabla \cdot \Pi + \nu \rho_{cr} (\uv_{cr}-\uv),
\label{eq:maxwell2}
\end{eqnarray}
where $\rho_{cr}$ is the mass density of the relativistic particles, and $\uv_{cr}-\uv$ is the drift speed of the cosmic rays relative to the plasma, which to zeroth order is equal to $\uv_s$. Since ${\bf j}_{th} = -{\bf j}_{cr}+c/(4\pi) \nabla \times \Bv$ the momentum equation (Eq.~\ref{eq:maxwell2}) has to satisfy:
\begin{eqnarray}
\rho \partial_t \uv &=&-\frac{{\bf j}_{cr} \times \Bv}{c}+\frac{1}{4\pi}(\nabla\times\Bv)\times\Bv- \nabla P -\nabla \cdot \Pi + \nu \rho_{cr} (\uv_{cr}-\uv).
\end{eqnarray}

In systems relevant for efficient diffusive shock acceleration, the upstream plasma can be considered cold and isotropic, such that $\nabla P=0$ and $\nabla \cdot \Pi = 0$. Additionally, the second term on the r.h.s. is much smaller than the other terms, and only contributes at very short wavelengths where the magnetic tension is sufficient to quench the instability. If we furthermore use that ${\bf j}_{cr}=n_{cr}e \uv_s =e \fv_1 c/3$, divide both sides by $\rho$, and write $n=n_{cr}/n_i$ the ratio between cosmic ray- and background nucleons, the momentum equation can be expressed in terms of $\fv_1$ as follows:
\begin{eqnarray}
\label{eq:u1v}
\partial_t  \uv =  -\frac{n e}{3 m_{p}}(\fv_1 \times \Bv) +\frac{c}{3}\nu n \fv_1.
\end{eqnarray}
For a linear analysis it suffices to look at the first order perturbation, for which the above can be expressed, using Eq.~\ref{eq:f1v}, as:
\begin{eqnarray}
\label{eq:u1vf2}
\partial_t  \uv_{(1)}&=&-\frac{nc}{3}\left(\partial_t \fv_{1(1)} + c\nabla f_{0(1)} + \frac{2}{5}c\nabla \cdot \fv_{2(1)}\right),
\end{eqnarray}
where the subscripts between brackets indicate unperturbed $(0)$ or perturbed $(1)$ variables.

Both of the two above equations are instances of the momentum conservation, viewed either through the forces ${\bf j} \times \Bv$ and frictional force $\rho \nu \uv$, or  through the pressure gradient and divergence. Feeding this into the induction equation, we can express the perturbed magnetic field in terms of $\fv_1$ and $\fv_2$:
\begin{eqnarray}
\label{eq:B1}
\partial_t^2 \Bv_{(1)} &=& \partial_t(\Bv_{(0)} \cdot \nabla)\uv_{(1)}=-\frac{nc}{3}\left(\Bv_{(0)} \cdot \nabla\right)\left(\partial_t \fv_{1(1)} + c \nabla f_{0(1)}+\frac{2}{5}c\nabla \cdot \fv_{2(1)}\right)
\end{eqnarray}
where we assumed a homogeneous background magnetic field, incompressibility, and $\uv_{(0)}=0$. For the rest of this section we consider a parallel shock in the $z$-direction, such that $\Bv_{(0)}=B_{(0)} \hat {\bf z}$, and we consider modes parallel to the original field, such that $k \cdot \Bv_{(0)}=0$.

\begin{figure}
\begin{center}
\includegraphics[width=\textwidth]{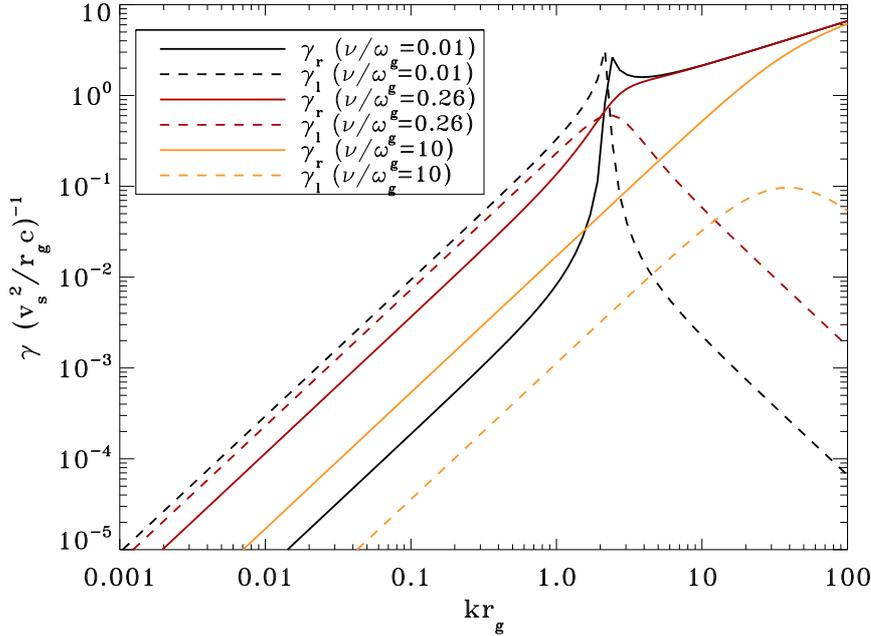}
\caption{
Growth rate of the tensor-mediated instability, from \citet{2011SchureBell}. The solid (dashed) lines indicate the right- (left-) hand mode. The different colours give the growth rates for different values of the effective collisionality of the cosmic rays, which aids in coupling the cosmic ray momentum equation to the momentum equation of the plasma. }
\label{fig:longwavelength}
\end{center}
\end{figure}

Equations~\ref{eq:f1v}--\ref{eq:B1} can then be combined \citep[see][]{2011SchureBell} to arrive at the dispersion relation: 
\begin{eqnarray}
\label{eq:dispf2wsmall}
\omega^2=
\pm\Omega^2\left(\frac{k^2 c^2}{5(3 \nu\mp i\omega_g)} - i \omega \right)
\left/
\left(\nu \mp i\omega_g+\frac{k^2 c^2}{5(3 \nu\mp i\omega_g)}\right)\right.,
\end{eqnarray}
where $\omega$ is the complex frequency, $k$ the wavenumber, $c$ speed of light, $\nu$ the effective scattering frequency, $\omega_g$ the gyrofrequency, and $\Omega=\sqrt{k j_{0} B_{(0)}/(\rho c)}$ contains the driving (return) current $j_0$ and is the growth rate of the non-resonant Bell instability. 
The upper signs correspond to the left-handed polarisation (which is the polarisation of a gyrating proton), and the lower signs to the right-hand polarisation. 
 This dispersion relation is valid in the linear regime as long as the Alfv\'enic stress due to $(\nabla \times \Bv)\times \Bv$ is small (for the parameters used to plot Figure~\ref{fig:longwavelength} beyond $kr_g\approx1000$). Also, since $\omega$ is always small compared to $k^2c^2/(5(3\nu \mp i \omega_g))$, it can in practise be ignored on the right hand side of Equation~\ref{eq:dispf2wsmall}.

The terms including factors of $k^2$ result from the inclusion of the stress tensor. Effects from the small-scale non-resonant instability are included through $\nu$, the effective scattering frequency. Scattering on small scales is expected to arise earlier than on long scales, since the growth rate increases rapidly for $kr_g \gg 1$.
Bohm diffusion is the regime where the effective scattering frequency is of the same order as the gyrofrequency, i.e. $\nu \approx \omega_g$.
We plot the growth rate as a function of wavenumber for different values of $\nu/\omega_g$ in Figure~\ref{fig:longwavelength}, for both the left-hand (dashed) and the right-hand polarisation \citep[for the full plots of the dispersion relation see the original paper: ][]{2011SchureBell}.

This method recovers the \citet{2004Bell} growth rate for the right-hand polarisation in the regime where $kc \gg \omega_g$. The resonant instability is only approximately captured in the mono-energetic approach (as can be seen from the peak around $k r_g \approx 2$ in Figure~\ref{fig:longwavelength}), which is why the familiar $k$-dependent growth rate around $k r_g=1$ is not present in this representation. Care should be taken when including the resonant instability to do the momentum integration with an appropriate upper limit for $p$, so as to not overestimate the growth rates on the long-wavelength end. It can be seen that both modes are unstable. Which of the polarisations dominates depends on the ration $\nu/\omega_g$; for $\nu/\omega_g < 1/\sqrt{3}$ the left-hand mode dominates, and vice versa for higher collisionality. When the collisionality is zero, $\nu=0$, only the left-hand mode is unstable and purely growing. The current-driven long-wavelength instability depends on the mediation of the short-scale instability through the stress-tensor and the `collisionality'.

\bigskip
Self-consistency between the equations for the cosmic rays and for the fluid are crucial in deriving the dispersion relation. In the following analysis we show that omission of the force due to friction in the momentum equation, as can often be done in other circumstances, would result in a completely different dispersion relation that shows a much more rapid growth and only declines at long wavelengths as $\sqrt{k}$. This is a result of an unbalanced frictional force, that results in the $\nu \fv_1$ term remaining in the momentum equation (Eq.~\ref{eq:u1vf2}) as a consequence of there not being a similar term in the momentum equation for the fluid.
This results in the following erroneous dispersion relation:
\begin{eqnarray}
\omega^2=\pm\Omega^2\left(\nu - i\omega+\frac{k^2 c^2}{5(3 \nu\mp i\omega_g)}\right)
/
\left(\nu - i\omega \mp i\omega_g+\frac{k^2 c^2}{5(3 \nu\mp i\omega_g)}\right)
\end{eqnarray}
The difference is only an additional $\nu$ in the nominator. For large $k$, the additional $\nu$ can be ignored such that on the short scales the result does not change. However, on the long-wavelength end the additional $\nu$ would change the result.
To lowest order in $k$ (thus ignoring the $k^2$ contribution from $\fv_2$), and in the limit $\omega \ll \omega_g$, the growth rate would change to:
\begin{eqnarray}
\gamma=\sqrt{\frac{\Omega^2\nu\left(\sqrt{\nu^2+\omega_g^2}\mp\nu\right)}{2(\nu^2+\omega_g^2)}}.
\end{eqnarray}

In the limit 
 where $\nu \ll \omega_g$, both modes would have similar growth rates, which would read:
\begin{eqnarray}
\gamma=\sqrt{\frac{\nu \omega_g}{2(\nu^2+\omega_g^2)}} \Omega = \sqrt{\frac{\nu}{2\omega_g}} \Omega = \sqrt{\frac{\nu k j_0 B_0}{2 \omega_g \rho c}}  \qquad (\nu \ll \omega_g).
\end{eqnarray}
It should be stressed this is not a physical solution.

\subsection{Ponderomotive instability}
\citet{2011Bykovetal} presented a long-wavelength instability that results from 
averaging the kinetic
equation for the relativistic particles, the equations of the bulk
plasma motions and the induction equation over the ensemble of the
short scale fluctuations produced by CR instabilities in the
collisionless regime e.g. by the fast Bell instability. 
To derive the growth rates of the modes in the long-wavelength regime $k \Lambda <1$, with $\Lambda=r_g/(\nu/\omega_g)$ the mean free path, the dispersion relation as in Eq.~\ref{dispers2asimpX0} derived from the collisionless kinetic equation approach is in this case not appropriate. 

In the
presence of the short-scale fluctuations, the momentum exchange
between the CRs and the flow in the hydrodynamic regime, results in
a ponderomotive force that depends on the CR current in the
mean-field momentum equation of bulk plasma \citep[][]{2011Bykovetal}. 
As a
result, there exist transverse growing modes with wavevectors along
the initial magnetic field with growth rates that are proportional
to the turbulent coefficients determined by the short scale
fluctuation:
\begin{eqnarray}
\gamma \approx \sqrt{\frac{\pi \sqrt{\langle {\bf b}^2\rangle}}{2 B_0} \frac{k j_0 \nu}{\rho c \omega_g}},
\label{eq:bykovgammalw}
\end{eqnarray}
where $\sqrt{\langle {\bf b}^2 \rangle}$ is the magnitude of the short-scale amplified magnetic field and
which holds for both polarisations.
The magnetic field amplification in that regime only
weakly depends on the shock velocity ($\gamma \tau \propto u_s^{-1/2}$ as follows from Eq.~(48) in
\citet{2011Bykovetal}, see also \citet{2011SchureBellb}], that is important for the evolution of the maximal
energy of CRs accelerated by DSA.
In the intermediate regime, $\nu/\omega_g < k r_g < 1$, the growth rate can be approximated as:
\begin{eqnarray}
\gamma=4\pi k \sqrt{\langle {\bf v}^2 \rangle},
\label{eq:bykovgammaim}
\end{eqnarray}
where $\sqrt{\langle {\bf v}^2 \rangle}$ is the amplitude of the short-scale turbulent bulk velocity.

The ponderomotive instability is a multi-layered phenomenon and the underlying physics is not immediately clear.  In the intermediate regime, the growth rate (Equation \ref{eq:bykovgammaim}) is independent of the CR current which suggests that it is not directly driven by CR streaming.  Instead, the growth time is equal to the time taken to cross a distance $1/k$ at the characteristic turbulence velocity.  This suggests that the magnetic field in this regime grows as a result of field-line stretching by already existing turbulent motions.  The long wavelength regime (Equation \ref{eq:bykovgammalw}) is complicated because it includes magnetic field on three different scales: {\em i}) $B_0$ on a scale comparable with or greater than the wavelength, {\em ii}) the magnetic field $\sqrt{\langle {\bf b}^2 \rangle}$ associated with the turbulence, and {\em iii}) the small scale magnetic field causing the scattering represented by the collision frequency $\nu$.  The challenge is to see why all three fields are important for the instability.  Is also important to check that the ${\bf j} \times \Bv$ momentum transfer between CR and the fluid is treated self-consistently in each case since errors in total momentum conservation can lead to an incorrect growth rate as shown in section 5.2.  In the next section we review the filamentation instability which is also due to the presence of turbulent magnetic field on scales smaller than $k r_g\sim 1$.  A further challenge is to ascertain whether there is any overlap in underlying physics between the ponderomotive and filamentation instabilities.

\subsection{Filamentation instability}
\label{sec:filamentation}

The expansion of the loops generated by the non-resonant instability on small scales can give rise to a further filamentation instability. Because the cosmic rays are focussed into filaments, the cosmic rays current locally increases. As a result magnetic fields around the loops further grow in strength, which again aids to focus the cosmic rays and increase the current. This was recently derived analytically and shown numerically by \citet{2011RevilleBell}.
The growth rate turns out to be independent of wavelength.
Again, the Vlasov equation is used to determine the distribution of cosmic rays. The local electric field can be expressed in terms of the vector potential such that:
\begin{eqnarray}
{\bf E}=u_s \nabla A_\parallel /c,
\end{eqnarray}
with $A_\parallel$ the magnitude of the vector potential parallel to the shock normal. Using this, the distribution function can be written as:
\begin{eqnarray}
\frac{\partial f}{\partial t} + c \frac{{\bf p}}{p} \cdot \nabla f + e\nabla(u_s A_\parallel /c) \cdot \frac{\partial f}{\partial {\bf p}} = 0.
\end{eqnarray}
Since it can be safely assumed that $\partial f/\partial p < 0$, the cosmic-ray number density is locally larger when $A_\parallel$ is positive and can be written as a number of position:
\begin{eqnarray}
n_{\rm cr} = n_0 + \frac{e u_s A_\parallel}{c^2} \int 8 \pi p f_0 {\rm d}p,
\end{eqnarray}
where $n_0=\int 4 \pi p^2 f_0 {\rm d}p$ and $f_0$ the isotropic part of the distribution function.
Using further that $j_{\rm cr}=n_{\rm cr} e u_s$ and the MHD equations, 
the evolution of the filamentation can be expressed in terms of the cosmic ray current as follows \citep[see][]{2011RevilleBell}:
\begin{eqnarray}
\frac{\partial^2 j_{\rm cr}}{\partial t^2} = \frac{e^2 n_0}{p_1} \frac{u_s^2}{c^2}\frac{B^2_\perp}{\rho c}j_{\rm cr} + \left(\left[(\uv \cdot \nabla)\uv\right] \cdot \nabla \right) j_{\rm cr} - (\uv \cdot \nabla) \frac{\partial j_{\rm cr}}{\partial t}.
\end{eqnarray}
The first term on the right-hand side is independent of the wavenumber and obviously dominates the other terms on long scales. Its growth rate is:
\begin{eqnarray}
\gamma=e \frac{u_s}{c} \sqrt{\frac{B_\perp^2}{\rho c} \frac{n_0}{p_1}} =\left(\frac{u_s}{c}\right)^2\left(\frac{U_{\rm cr}}{\rho u_s^2}\right)^{1/2}\frac{e B_\perp}{p_1\sqrt{\ln(p_2/p_1)}},
\end{eqnarray}
for a cosmic ray spectrum with a power law slope $q=4$ with a given minimum ($p_1$)and maximum ($p_2$) momentum cut-off. The requirement on the cosmic rays driving the instability is that they are not trapped within the cavities, such that $p_1 c \gg e A_\parallel u_s/c$, with $A_\parallel$ the vector potential. This condition was also assumed when deriving the instability. The growth rate is linearly dependent on the value of the amplified small-scale magnetic field and decreases upstream of the shock when $p_1$ increases and $U_{\rm cr}$ decreases. 

When compared to the non-resonant growth rate on small scales, the growth rates are equal when 
\begin{eqnarray}
k r_g = \frac{u_s}{c}\frac{\langle B_\perp^2 \rangle}{B_0^2}.
\end{eqnarray}
Non-linear simulations of the non-resonant instability indicate that the amplified field can reach values of $~30 B_0$ \citep{2004Bell, 2009RiquelmeSpitkovsky}, such that the above condition can be satisfied for $k r_g > 1$. Filling in numbers comparable to those used in the previous sections, the instability operates provided that:
\begin{eqnarray}
E_{\rm cr} \geq \left(\frac{B_\perp}{100~\mu G}\right) \left(\frac{k_{\rm max}}{2 \times 10^{15}~{\rm cm}}\right)^{-1}\left(\frac{u_s}{10^9~{\rm cm/s}}\right)~{\rm TeV}.
\label{eq:eminfil}
\end{eqnarray}
When compared to the growth rate of the long-wavelength instability, they become comparable when:
\begin{eqnarray}
k r_g = \left( \frac{5 u_s}{c}\frac{\langle B_\perp^2 \rangle}{B_0^2}\right)^{1/3}.
\end{eqnarray}
Thus, for the longest wavelengths, the filamentation instability may dominate if the small-scale field is sufficiently amplified and as long as condition \ref{eq:eminfil} is satisfied.

\subsection{Firehose instability}
The short-wavelength non-resonant instability in Section~\ref{sec:nonresonantbell} was driven by the current, i.e. the first order anisotropy in the cosmic ray distribution. Asymmetry in the second order anisotropy, the pressure tensor, can drive the firehose or mirror instability.  It is distinctly different from the long-wavelength instability that was discussed in Section~\ref{sec:stresstensor}: that one was driven by the current, and the stress-tensor (off-diagonal terms of the pressure tensor) and small-scale collisions mediate this driving source to cause instability on long length-scales and the opposite (left-hand) polarisation. 

Again, we will have to consider the distribution function up to second order anisotropy, which we can write in the form
\begin{equation}\label{distrF0}
f({\bf p})=\frac{N}{4\pi}\left[1+3\frac{u_s}{c}\mu+\frac{\delta}{2}\left(3\mu^{2}-1\right)\right],
\end{equation}
where  $\theta$ - particle pitch-angle, $\mu=\cos\theta$ ,
$\delta(p)$ - is the magnitude of the second harmonic anisotropy, which is normally of the order of $u_s^2/c^2$. 
Indeed, it is the second harmonic anisotropy that constitutes the source of the CR-firehose instability on the magnetic field
amplification. It is instructive to summarize the  growth rates for magnetic
instabilities that the quasi-linear theory predicts for weakly
anisotropic CR distributions of the above form.
Following the standard
linear analysis of the kinetic equation in the intermediate
regime $r_g/\Lambda < x_0< 1$ \citep[see e.g.][]{2011Bykovetal}, with $\Lambda$ again being the mean free path, one may get
the following dispersion relation if collisions can be neglected:
\begin{equation}\label{dispers}
\frac{\omega^{2}}{v_{a}^{2}k^{2}}=\left[1\mp
\frac{k_{0}}{k}\left\{(A_{0}-1)+\frac{\delta c}{u_{s}}A_{1}\right\}\right].
\end{equation}
\begin{equation}\label{koeffA}
A_{0,1}\left(x_1,x_2\right)=\int_{p_1}^{p_2}\sigma_{0,1}\left(p\right)N\left(p\right)p^{2}dp
\end{equation}
\begin{equation}\label{sigma0Int}
\sigma_{0}\left(p\right)=\frac{3}{4}\int_{-1}^{1}\frac{\left(1-\mu^{2}\right)}{1\mp
x\mu}d\mu,
\end{equation}
\begin{equation}\label{sigma1Int}
\sigma_{1}\left(p\right)=\frac{3}{4}\int_{-1}^{1}\frac{\left(1-\mu^{2}\right)\mu}{1\mp
x\mu}d\mu
\end{equation}
where $\displaystyle
k_{0}=\frac{4\pi}{c}\frac{en_{cr}u_{s}}{B_{0}}$, $x=k r_{g}(p)$ ,
$x_1=k r_{g}(p_1)$ , $x_2= k r_{g}(p_2)$. The signs $\pm$
correspond to the two opposite circularly polarized modes under
investigation. 
The second term on the right hand side is the instability that is driven by the cosmic ray current, and has a non-resonant form \citep[][see previous section]{2004Bell} when $k r_g$ is large, and a resonant form around $k r_g$ of the order $(1)$ \citep[e.g.][and the references therein]{1981Achterberg,2003Zweibel,2006Pelletieretal,2006Marcowithetal,2009AmatoBlasi}.
The last term in
the r.h.s. of Eq.~\ref{dispers} represents the CR firehose
instability. In the long-wavelength regime  $x_{m} \ll 1$  a
simplified form of Eq.~\ref{dispers} can be derived
\begin{equation}\label{dispers2asimpX0}
\frac{\omega^{2}}{v_{a}^{2}k^{2}}=1\mp
\frac{k_{0}r_{g0}}{5}\left[x_2\pm\frac{\delta c/u_s\ln(p_2/p_1)}{\left(1-(p_2/p_1)^{-1}\right)}\right].
\end{equation}
The growth rate of the firehose instability due to the CR pressure anisotropy is found in the last term of Eq.~\ref{dispers2asimpX0}. It requires that the parallel pressure exceeds the perpendicular pressure such that $P_\parallel > P_\perp+B^2/(4\pi)$  \citep[for more details see e.g.~][]{2011Bykovetalreview}.

\subsection{Instabilities driven by the cosmic ray pressure gradient in the shock precursor}

On scales large compared to the gyro-radius (i.e. scales where the driving particles are strongly magnetised) the action of the cosmic rays can be simplified to a bulk force on the background plasma which is just given by the gradient of the cosmic ray pressure,
\begin{equation}
P_{cr}= \int {pv\over 3} 4\pi p^2 f(p) {\rm d}p
\end{equation}
This  follows naturally from considerations of momentum balance, the above pressure integral being just the flux of momentum associated with the cosmic rays. 

The interesting thing about this bulk force is that it is not related in any simple way to the local density either of mass or of scattering centres (this is a consequence of the collective nature of the electro-magnetic interactions in the plasma).  A gravitational field, by contrast, would produce a force that is always strictly proportional to the local mass density (this is just Einstein's equivalence principle).  Similarly a flux of particles interacting by two-body scatterings would produce a force that is proportional to the density of scattering centres, radiation pressure resulting from Thompson scattering on electrons being an example.  No such simple relation holds for the cosmic ray scattering, which is a complicated function of the power-spectrum of structure in the magnetic field on scales comparable to the particle gyro-radius and, to lowest order, can be calculated using the methods of quasi-linear theory. For our purposes it is enough to just assume that there is some effective diffusion coefficient $\kappa(p)$ and that on the scales of interest to us the cosmic ray transport can be represented as a simple diffusion process with a flux proportional to the local gradient,
\begin{equation}
{\partial f\over\partial t} + u\cdot\nabla f = {1\over 3}\nabla\cdot u\,p {\partial f\over\partial p}  + \nabla\cdot(\kappa \nabla f)
\end{equation}
where $\kappa$ is the diffusion tensor (often approximated as a scalar diffusion coefficient).

A given element of background plasma, of local density $\rho$ experiences an acceleration
\begin{equation}
\frac{\partial u}{\partial t}=-{1\over\rho}\nabla P_{cr}
\end{equation}
and thus, even if the cosmic ray pressure is very uniform, local small-scale variations in density induce acceleration fluctuations which in turn lead to velocity fluctuations which can feed back into density fluctuations.  The usual approach to studying instabiities, where one assumes a uniform steady background and then does a Fourier analysis of the modes, fails in this case because the non-stationary and non-uniform nature of the shock precursor region is the ultimate source of free energy driving the instability.  Indeed the very question of how to define an instability in such a system is an interesting one with no obvious answer.
 
In \citet{1986DruryFalle} a solution to these problems was developed based on a two-scale expansion of the governing equations.  In this approach one looks at  small wave-length high-frequency modes propagating on a smoothly varying background.   In the absence of the cosmic ray pressure the basic modes are then just sound waves and the wave amplitude satisfies a conservation equation for the wave action (this follows from Noether's theorem because in the high-frequency limit the precise phase of the wave becomes unimportant, and there is thus an asymptotic symmetry related to the arbitrariness of the phase angle).   In the case of one spatial dimension this is
\begin{equation}
{\partial {\cal A}\over \partial t} +{\partial\over\partial z}\left[ \left(u\pm c_s\right)\cal A\right]
= 0
\end{equation}
where the wave action $\cal A$ is the acoustic wave energy density divided by the local co-moving frequency
, $c_s$ is the sound speed and the sign of $\pm$ corresponds to left- and right-travelling modes.

Including cosmic ray effects in the two-fluid approximation it is then possible to show \citep[see][]{1986DruryFalle} that the wave action equation acquires a non-zero right hand side,
\begin{equation}
{\partial{ \cal A}\over \partial t} +{\partial\over\partial z}\left[ \left(u\pm c_s\right)\cal A\right]
= {\cal A} \left[ -{\gamma_{cr} P_{cr}\over\rho \kappa}
\pm {\partial P_{cr}/\partial z\over c_s\rho} \left(1 +  \frac{\partial \ln\kappa}{\partial\ln\rho}\right)\right]
\end{equation}
where $\gamma_{cr}$ is an effective adiabatic index for the cosmic rays and $\partial \ln\kappa/\partial\ln\rho$ is the extent to which fluctuations in density induce fluctuations in the diffusion coefficient.

The first term is a linear damping term related to the cosmic ray diffusion, as derived earlier by \citet{1981Ptuskin}.  More interesting for our purpose is the second term which is a potentially de-stabilising term related to the cosmic ray pressure gradient.   By formulating the problem in this way it is possible to clearly separate out the conservative effects of the changing background, encapsulated in the conservation of wave action, from the non-conservative effects of the cosmic ray pressure. This gives a precise instability criterion,
\begin{equation}
\frac{1}{\rho c_s}\frac{\partial P_{cr}}{\partial z} \left(1 +  \frac{\partial \ln\kappa}{\partial\ln\rho}\right)> {\gamma_{cr} P_{cr}\over \rho\kappa}
 \end{equation}
A gradient in the cosmic ray pressure can arise when the cosmic rays make up a significant fraction of the total pressure, and is of interest in the shock precursor region of an efficiently accelerating shock wave. 

If we introduce a length-scale for the cosmic ray pressure,
\begin{equation}
{\partial P_{cr}\over\partial z} = {P_{cr}\over L}
\end{equation}
the condition for instability can be written as
\begin{equation}
L < {\kappa\over \gamma_{cr} c_s}\left(1+ \frac{\partial \ln\kappa}{\partial\ln\rho}\right).
\end{equation}
Noting that in a shock precursor $L\sim \kappa/u_s$ where $u_s$ is the shock velocity, it is clear that the shock precursor region will be generically unstable at high shock Mach numbers unless $ \partial \ln\kappa/\partial\ln\rho$ is identically $-1$.  This is confirmed by numerical simulations of modified shocks in the two-fluid approximation which exhibit instabilities unless the product $\rho\kappa$ is artificially kept constant. Indeed it was this phenomenon in the early calculations of Dorfi which led to the discovery of the instability.    It should be noted that the instability only operates as a fluid element is advocated through the precursor, so the maximum growth is limited to an amount of order $\exp(M)$ which can however be very large for high Mach numbers.

In addition to the acoustic modes there are also non-propagating entropy modes.  The entropy of each fluid element is conserved and therefore the entropy modes can not grow, at least until secondary shocks form. The modes do couple to acoustic modes which can then be amplified.  Such entropy modes are perhaps better thought of as dense clumps in pressure equilibrium with their surroundings and the differential acceleration forces resulting from the cosmic ray pressure will set these in motion relative to their less dense surroundings.  This motion will then be transmitted to the surroundings in the form of acoustic waves which, when propagating parallel to the shock normal, will be amplified unless the diffusion coefficient scales inversely with the density according to the above analysis.  If we now consider the very special case of a sinusoidal density perturbation with wave vector perpendicular to the shock normal, it is clear that it too will induce motions unless the diffusion is constant and independent of density.  Thus in three dimensions it is impossible to fully stabilize a clumpy shock precursor.  The condition for parallel stability implies transverse instability and vice-versa.  

It is also worth noting, as pointed out by \citet{2007GiacaloneJokipii}, that even with no precursor effects a clumpy upstream medium will induce strong post-shock vorticity and down-stream magnetic field amplification.  The above analysis indicates that similar processes can work upstream if there is a strong cosmic-ray precursor and thus produce magnetic field amplification by what is in essence a bulk hydrodynamic effect. The scales on which this takes place can be large compared to the particle gyro-radii and is determined by the characteristic length-scales of the initial density fluctuations.

\section{Deviations from the ``-2'' power law index}
\label{sec:nonideal}
\subsection{The spectral index at oblique shocks}
\label{sec:oblique}

The relative motion of upstream and downstream scatterers 
imparts energy to CR as they bounce back and forth across a shock.
By Lorentz transformation, the mean fractional
energy gain is $u_s/c$ on each passage from upstream to downstream and
back to upstream for a strong non-relativistic shock.
As discussed in Section~\ref{sec:dsa}, the resulting CR spectrum is determined by the statistical distribution
of the number of times a CR crosses the shock before escaping downstream.
CR cross the shock at a rate $n_sc/4$ where $n_s$ is the 
CR number density at the shock.
This is balanced by the rate $n_\infty u_s/4$ at which CR advect away
downstream of the shock, where $n_\infty$ is the CR number density
far downstream.  Hence the fractional number of CR lost after each crossing is
$(n_\infty/n_s)(u_s/c)$.
In the limit of small shock velocity, the VFP equation dictates that $n_\infty=n_s$
and the fractional number lost is $u_s/c$.
When combined with the fractional energy gain of $u_s/c$, a
power law CR distribution results with differential energy spectrum 
$n(E)dE\propto E^{-\gamma}dE$ where $\gamma=2$.  If, as we show below,
$n_s$ can differ from $n_\infty$, 
the spectral index is 
\begin{eqnarray}
\gamma =1+n_\infty/n_s.
\label{eq:idealindex} 
\end{eqnarray}

The above argument predicts a universal $E^{-2}$ spectrum for
all high Mach number non-relativistic shocks, 
whether they are perpendicular, parallel or oblique.
However, the argument depends on the result that $n_\infty=n_s$.
It has been clear for a while that this breaks down when the shock
velocity is relativistic \citep[e.g.~][ and Spitkovsky 2011 these proceedings]{2001Achterbergetal}, 
but it has recently been shown (Bell et al 2011) that
departures from this spectrum occur at
shock velocities as low as $\sim c/30$ 
as found in young supernova remnants (SNR).
The departure is particularly strong for shocks that are nearly perpendicular
and when the CR mean free path $\Lambda$ is larger than the Larmor radius $r_g$.
As shown in the previous section, 
the precursor scaleheight at a perpendicular shock 
is $ L\sim (r_g/\Lambda)(c/u_s) r_g$.
According to this formula, the upstream scaleheight can be very short
causing a discontinuity in the CR density gradient across the shock.
Kinetic theory does not allow discontinuities
occur over distances less than a CR Larmor radius since
CR gyration imposes a smoothing distance of a Larmor radius on the
CR distribution function.
If the precursor scaleheight is not much larger than the Larmor radius,
the overall CR change in density across the shock takes place partly
downstream as well as upstream. The result is that the CR density $n_s$ at the shock
is less than the density far downstream $n_\infty$, the spectral index
$\gamma$ is greater than two, and the CR spectrum is steepened
as indicated by equation \ref{eq:idealindex}.

In contrast, solution of the VFP equation shows that the spectrum is flattened
if the shock is oblique and more than $10-20^\circ$ from perpendicular, depending on the shock velocity.
Compression at the shock increases the perpendicular component of the magnetic field because it is 
frozen in to the background plasma, whereas the parallel component of the field
is unchanged by the shock.
Consequently the magnetic field increases in magitude and 
changes direction at an oblique shock.
The change in field acts as a partial magnetic mirror 
which reflects back upstream some CR trying to cross the shock
into the downstream plasma.
The shock acts as a partial snowplough, pushing CR ahead of it.
This produces a local excess in the CR number density at the shock
such that $n_s> n_\infty$, and the CR spectrum is flattened
in accordance with equation \ref{eq:idealindex} 
 giving $\gamma <2$.

Clearly, the universal strong shock spectrum, $\gamma =2$, does not hold 
for young SNR shocks. This is is consistent with radio observations which
exhibit significantly steepened spectra in very young SNR expanding at
high velocity \citep{2011Belletal}.
The effect is not confined to high velocity shocks.
The crucial parameter is the ratio of the shock velocity to the velocity
of the accelerating particle, so the spectra of sub-relativistic particles accelerated by heliospheric shocks
may be expected to show a related departure from $\gamma = 2$.

\subsection{Non-linearity and time-dependence}

From observations only direct evidence of acceleration of electrons in supernova remnant blast wave exists through the observation of narrow synchrotron rims in X-ray and radio-synchrotron emission from a more extended region. In theory there is no reason why protons and heavier nuclei would not be accelerated through the same process. Gamma-ray observations are currently hinting towards observations of escaping nuclei interacting with molecular clouds, and perhaps also in situ protons at SNR shock waves. Protons could potentially be much more abundant (estimates of a factor 1000 are not uncommon), due to easier injection and less radiation losses, and may reach higher energies. Efficient proton acceleration would therefore have important consequences on the shock structure and temperature.  The non-linear effects due to cosmic rays constituting a significant fraction of the energy at the shock have been widely discussed and modelled \citep{1995Ellisonetal,1997Malkov,2002Blasi,2005KangJones,2006AmatoBlasi,2008Vladimirovetal, 2009Kangetal,2009Patnaudeetal, 2010Capriolietal, 2010Ferrandetal}. Especially towards the higher-energy end of the spectrum, the spectral index can significantly flatten due to the higher overall compression ratio that is probed by the more energetic cosmic rays. On the low-energy end spectral steepening can occur, although this is only important in low-Mach number shocks unless almost all of the energy goes into cosmic rays \citep[e.g.][]{2010Vinketal}. It may be important to consider in simulations of clusters of galaxies, in which low-Mach number shocks appear to accelerate particles \citep{2003Ryuetal,2006Pfrommeretal,2009Vazzaetal}. The transfer of energy to cosmic rays may further be seen in temperature deviations behind the shock, where part of the energy that was supposed to heat the plasma has effectively gone into a cosmic ray component \citep{2009Helderetal, 2009Patnaudeetal}.

Whether a shock can be an efficient accelerator, and what the resulting cosmic ray spectrum looks like, is also dependent on the environment the shock is running into. A core-collapse SNR will have a different evolution of the shock velocity from a SNR evolving in a homogeneous medium as may be the case for most Type Ia SNe. The time-dependent evolution will determine the cumulative spectrum \citep{2010Schureetal}. The environment may in some cases also affect the damping through ion-neutral collisions \citep{2007Revilleetal}, the detectable emission through various energy-exchange processes \citep[e.g.][]{2011Raymondetal} or through surrounding moldecular clouds enhancing the target density for pion creation from escaping cosmic ray protons \citep[e.g.][]{2011Ohiraetal}  

Apart from magnetic field amplification, progress has been booked on the theory of DSA that deviates from the ideal case. The powerlaw of the cosmic rays in reality is a complex addition of time-dependence, shock obliquity and shock speed, as has been discussed in Section~\ref{sec:oblique}. The theory on how cosmic rays may escape upstream is also under active development, with recent papers by \citet{2011Drury, 2011OhiraIoka}

\section{Discussion and conclusion}
\label{sec:discussion}
The importance of magnetic field amplification has been recognised since the early developments of the theory of DSA, and in the past decade significant progress has been made in this field. Driven by observations that strongly indicated amplification of factors 10-100, various theories have been developoed to explain this intrinsically nonlinear process. A distinction between scales shorter and longer than the gyroradius of the driving cosmic rays is made. Most rapid amplification can be achieved on short scales, but amplification on longer scales is paramount in accelerating cosmic rays to the PeV energies that they are believed to gain in galactic sources. Currently, SNRs still seem the best candidate, but the process is independent of the type of shock wave and other sources may well contribute. The nonlinear behaviour of the various instabilities remains an area of active research and more work is required to satisfactorily decide in which regimes various instabilites operate and dominate. Also the saturation level needs to be determined. Observations that determine more precisely direction, degree of polarisation, on different length and time scales could aid in constraining the theory. The high-energy end of the emission spectrum can be better probed with the current observatories that operate over a range of wavelengths in the gamma-ray regime. Most notably the low-energy end as probed by Fermi-LAT, as well as Cherenkov telescopes for the high-energy end, such as HESS. The next generation telescopes, such as CTA, will certainly be extremely useful, if the observed energy is pushed up high enough to really distinguish between electron- and proton- based emission processes. High-resolution radio measurements of magnetic field strength, and polarimetry in the X-ray band, are other items on the wish list.

\begin{acknowledgements}
We would like to acknowledge ISSI for their support during, and the organisation of, the workshop on particle acceleration in cosmic plasmas. 
K.M.S. and A.R.B. acknowledge support from the UK Science Technology and Facilities Council grant ST/H001948/1; and from the European Research Council under the European Community's Seventh Framework Programma (FP7/2007-2013) / ERC grant agreement no. 247039.
A.M.B was supported in part by the RAS Programs, by the RFBR grant
11-02-12082-ofi-m-2011, and also by the Russian government grant
11.G34.31.0001 to the Saint-Petersburg State Politechnical
University.
\end{acknowledgements}

\bibliographystyle{aps-nameyear}      
\bibliography{../adssample}   

\end{document}